\documentclass[manuscript, sigplan,screen]{acmart}

\setcopyright{none}
\renewcommand\footnotetextcopyrightpermission[1]{}

\usepackage{xspace}
\usepackage{enumitem}
\usepackage{graphicx}
\usepackage{subcaption}
\usepackage{caption}
\usepackage{tikz}
\usepackage{listings}
\AtBeginDocument{%
  }

\setcopyright{acmlicensed}
\copyrightyear{2018}
\acmYear{2018}
\acmDOI{XXXXXXX.XXXXXXX}
\acmConference[Dafny Workshop, POPL '26]{}{11th January, 2026}{
Rennes, France}

\usepackage{xcolor}
\usepackage{listings}

\definecolor{dafnypurple}{RGB}{136,0,136}
\definecolor{dafnykeyword}{RGB}{153,102,0}
\definecolor{dafnycomment}{RGB}{0,128,0}
\definecolor{dafnynumber}{RGB}{0,102,204}
\definecolor{dafnygreen}{RGB}{0,150,0}
\definecolor{posHighlight}{RGB}{0,110,0}
\definecolor{highlightred}{RGB}{255,0,0}
\definecolor{bordergray}{RGB}{180,180,180}

\lstdefinelanguage{Dafny}{
  morekeywords={
    method,var,while,if,else,ensures,requires,invariant,returns,assert,
    true,false,forall,exists,match,case,ghost,function,class,new,type,trait
  },
  sensitive=true,
  morecomment=[l]{//},
  morecomment=[s]{/*}{*/},
  morestring=[b]",
}

\usepackage[ruled,vlined,linesnumbered]{algorithm2e}
\usepackage[textsize=scriptsize]{todonotes}

\newcommand{\ToolName}{DafnyPro\xspace}

\begin{document}

\title{\ToolName: LLM-Assisted Automated Verification for Dafny Programs}

\author{Debangshu Banerjee\footnotemark[2]}
\affiliation{%
  \institution{UIUC}
  \city{Champaign, IL}
  \country{USA}
}
\email{db21@illinois.edu}

\author{Olivier Bouissou}
\affiliation{%
  \institution{Amazon Web Services}
  \city{Boston, MA}
  \country{US}}
\email{obouisso@amazon.com}

\author{Stefan Zetzsche}
\affiliation{%
  \institution{Amazon Web Services}
  \city{London}
  \country{UK}}
\email{stefanze@amazon.co.uk}

\renewcommand{\shortauthors}{Banerjee et al.}

\begin{abstract}
We present DafnyPro, an inference-time framework that enhances LLMs for generating verification annotations in Dafny. DafnyPro comprises three key components: a diff-checker that prevents modifications to base program logic, a pruner that removes unnecessary invariants, and a hint-augmentation system that retrieves and applies predefined, problem-independent proof strategies. We evaluate DafnyPro using Claude Sonnet 3.5 and 3.7 on four benchmarks: Clover, MBPP-Dafny, HumanEval-Dafny, and DafnyBench, achieving consistent performance gains in all cases. Notably, on DafnyBench, the most challenging benchmark, Claude Sonnet 3.5 enhanced with DafnyPro achieves 86\% correct proofs, a 16 pp improvement over the base model. We also fine-tune two Qwen models on training data derived from verification attempts by larger models enhanced with DafnyPro. Our 7B and 14B models achieve 68\% and 70\% correct proofs on DafnyBench, respectively, demonstrating that smaller models can maintain high verification accuracy.
\end{abstract}


\maketitle

\footnotetext[2]{The author worked on this project during their internship at Amazon Web Services in Boston, MA, US.}

\section{Introduction}

\begin{figure}[t] % h = here, t = top, b = bottom, p = page of floats
    \centering
    \includegraphics[width=\columnwidth]{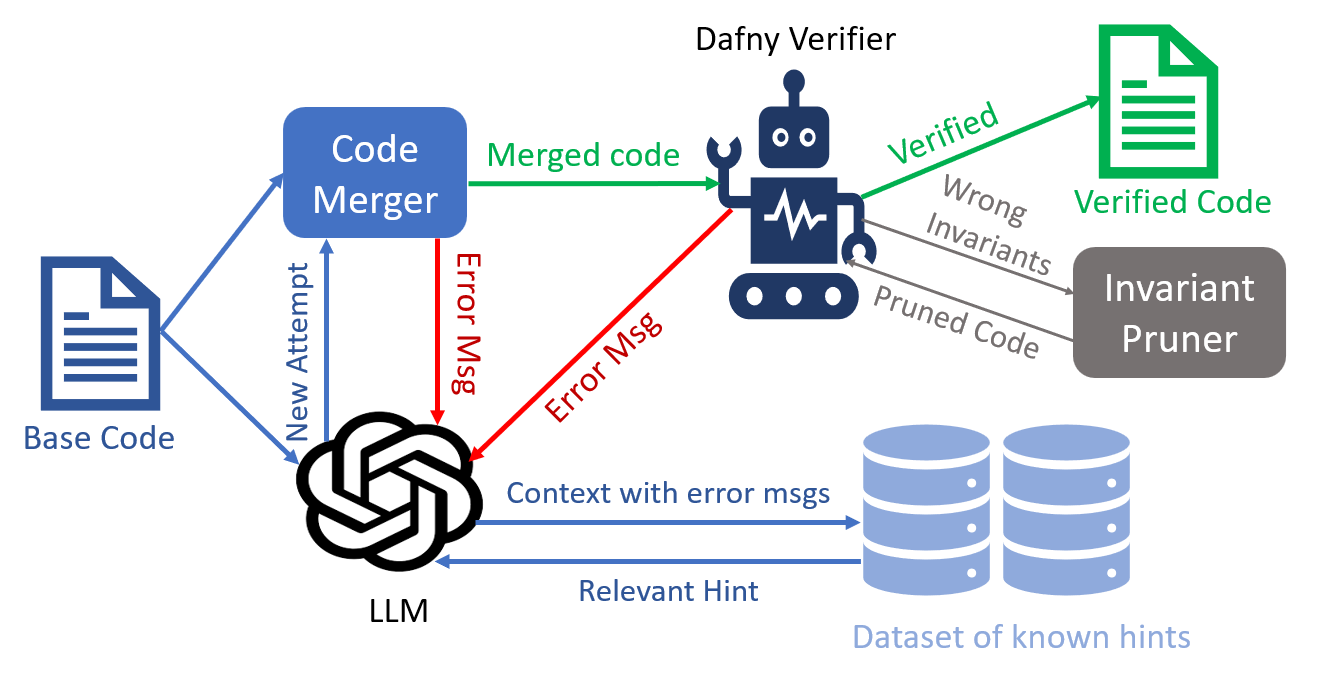} % replace with your filename
    \caption{Overview of the DafnyPro pipeline: (a) Merging, (b) Pruning, (c) Hint-Augmentation.}
\label{fig:DafnyProPipeline}
\end{figure}

Formal verification provides mathematical proofs that programs satisfy their specifications, offering stronger guaranties than testing or fuzzing. However, verification requires domain expertise to generate appropriate annotations—such as loop invariants and ranking functions—that enable automated verification tools. Writing these annotations manually is tedious and time-consuming.

Recent advances in large language models (LLMs) have enabled automating this process. Existing approaches \cite{dafnybench,clover,mbppDfy} iteratively generate annotated programs and use verifier feedback to guide subsequent generations. However, these frameworks suffer from two critical limitations: (a) LLMs sometimes alter the base code logic to pass verification without producing correct annotations (Fig \ref{fig:exCheating}), and (b) they fail to leverage known proof strategies that have proven effective in related problems (Fig \ref{fig:hintAug}).

In this paper, we introduce \textbf{\ToolName}, an inference-time framework for generating verification annotations in Dafny \cite{dafny}. \ToolName addresses these limitations through three mechanisms: (i) using Dafny's parser to reject any modifications to base code logic, (ii) iteratively refining annotations by adding or pruning based on verifier feedback, and (iii) retrieving and applying predefined, problem-independent proof strategies during generation.

\ToolName achieves 86\% correct proofs on DafnyBench with Claude 3.5 Sonnet \cite{claude35sonnet}, a 16 percentage point improvement over prior state of the art. We also introduce fine-tuned models for efficient deployment: our 7B model achieves 68\% and our 14B model achieves 70\% and on DafnyBench.

Our main contributions are as follows:
\begin{itemize}[leftmargin=*]
\item \ToolName: an inference-time framework achieving state-of-the-art accuracy on DafnyBench and other benchmarks.
\item Fine-tuned models (7B and 14B) enabling efficiently deployable verification with competitive performance.
\end{itemize}

\section{Related Work}

\paragraph{Inference-time methods}
Several works \cite{dafnybench, clover, mbppDfy, laurel} improve formal verification performance through inference-time strategies without model retraining.
DafnyBench \cite{dafnybench} provides a comprehensive benchmark and standardized setup for evaluating LLM verification capabilities.
Both Clover \cite{clover} and DafnyBench \cite{dafnybench} introduce multi-turn procedures that iteratively generate annotations, invoke the Dafny verifier, and use feedback from failed attempts to guide refinement.
Laurel \cite{laurel} extends this approach by automatically generating missing assertions using LLMs.
However, these methods do not prevent models from altering base code logic or systematically leverage known proof strategies—limitations that \ToolName addresses.

\paragraph{Model fine-tuning}
Complementary work focuses on adapting model weights to improve annotation generation.
Dafny-Annotator \cite{dafnyAnnotator} performs supervised fine-tuning on DafnySynth, a synthetic dataset derived from DafnyBench.
PREFACE \cite{preface} applies preference-based reinforcement learning with reward signals from the formal verifier to iteratively improve verification success.
Our fine-tuned models build on these insights while maintaining resource efficiency.

\section{Technical Contribution}
\label{sec:techContribution}

\ToolName takes an unannotated base program with its specifications and uses an LLM to generate annotations that prove correctness with respect to those specifications.
We identify three key limitations in existing approaches and propose targeted solutions that yield significant improvements across multiple datasets.
Below, we describe each solution in detail.

\subsection{Preventing Code Modifications}

\begin{figure}
    \centering
\begin{minipage}{0.96\columnwidth}
\begin{lstlisting}
(*@\setlength{\fboxsep}{0pt}\colorbox{highlightred}{\strut \color{white} b := a;  // Base code altered (was: b := a+1)}@*)
while ( b < n )
  invariant a <= b <= n
{
  if ( X[b] <= p ) {
    var t := X[b];
    X[b] := X[a];
    X[a] := t;
    a := a + 1;
  }
  b := b + 1;
}
\end{lstlisting}
\end{minipage}
\caption{LLM modifies base code instead of only adding annotations, compromising soundness.}
\label{fig:exCheating}
\end{figure}

Existing works \cite{dafnybench, clover} regenerate the entire Dafny program with annotations rather than generating annotations separately, since the latter is difficult when loops appear in multiple locations, are nested, or contain assertions at various points. However, regenerating the entire program risks altering its logic—for example, by changing variable initialization as shown in Fig.~\ref{fig:exCheating} (base code in Appendix~\ref{appen:diffCheck}). Such modifications, often called "cheating," are not reliably detected by the ad hoc checkers used in \cite{dafnybench}, which miss corner cases and compromise soundness. Similar behavior has been observed in coding and theorem-proving tasks, termed "reward hacking" or "reward tampering" \cite{rewardHacking, rewardHackingDefinition}.

Our diff-checker takes a principled approach: it uses the Dafny parser to extract added annotations (e.g., invariants, assertions) and verifies that the regenerated code, once stripped of annotations, is identical to the original base code. This guarantees soundness and provides a systematic method for parsing annotations, useful for data collection. For instance, existing datasets such as DafnyBench \cite{dafnybench} could leverage it to more accurately remove annotations and correct erroneous base codes (see Section~\ref{sec:dafnyproEval}).

\subsection{Removing Unnecessary Clauses}

\begin{figure}[t]
\centering
\begin{minipage}{0.96\columnwidth}
\begin{lstlisting}
method onlineMax(a: array<int>, x: int) returns (ghost m: int, p: int)
  requires 1 <= x < a.Length
  requires a.Length != 0
  ensures x <= p < a.Length
  ensures forall i :: 0 <= i < x ==> a[i] <= m
  ensures exists i :: 0 <= i < x && a[i] == m
  ensures x <= p < a.Length-1 ==> (forall i :: 0 <= i < p ==> a[i] < a[p])
  ensures (forall i :: x <= i < a.Length && a[i] <= m) ==> p == a.Length - 1
{
  p := 0; var best := a[0]; var i := 1; i := x;
  // ... Code Block
  while i < a.Length
    invariant x <= i <= a.Length
    invariant forall j :: 0 <= j < x ==> a[j] <= m
    invariant exists j :: 0 <= j < x && a[j] == m
    invariant forall j :: x <= j < i ==> a[j] <= m
    invariant i < a.Length ==> p == 0
    (*@\setlength{\fboxsep}{0pt}\colorbox{posHighlight}{\strut \color{white} // Pruned: non-inductive and unnecessary} @*)
    (*@\setlength{\fboxsep}{0pt}\colorbox{posHighlight}{\strut \color{white} // invariant i == a.Length ==> p == a.Length - 1} @*) 
  {
    if a[i] > best {
      p := i;
      return;
    }
    i := i + 1;
  }
  p := a.Length - 1;
}
\end{lstlisting}
\end{minipage}

\caption{Pruning unnecessary invariants enables verification, avoiding wasted feedback iterations.}
\label{fig:pruningBenefit}
\end{figure}

Fig.~\ref{fig:pruningBenefit} illustrates a second limitation: LLMs may introduce unnecessary invariant clauses that are not required to prove the post-condition (details in Appendix~\ref{appen:pruningDetails}). When these unnecessary invariants cannot be proven inductively, the feedback-based iterative generation used in prior work \cite{clover, dafnybench} wastes attempts fixing such clauses. To address this, we introduce a greedy invariant clause pruning strategy.

For each LLM-generated annotated program, the pruner uses the Dafny verifier to identify all non-inductive invariant clauses and removes them. It then attempts to verify the post-condition with the remaining clauses. This process continues until either (i) the post-condition is proved, or (ii) the verifier reports that the post-condition cannot be proved with the current clauses. If pruning fails, the pruner restores the original annotated version and proceeds to the next generation attempt. This approach removes unnecessary clauses while preserving the possibility of correcting non-inductive but essential clauses in subsequent iterations.

\subsection{Incorporating Domain Knowledge}

\begin{figure}[t]
    \centering
\begin{minipage}{\columnwidth}
\begin{lstlisting}
method   <T(==)>(a: array<T>, key: T) returns (b: bool)
  ensures (multiset(a[..])[key] == 1) <==> b
{
  var i := 0; b := false; var keyCount := 0;
  while i < a.Length
    invariant 0 <= i <= a.Length
    invariant keyCount == multiset(a[..i])[key]
    invariant b <==> keyCount == 1
  {
    if (a[i] == key){
      keyCount := keyCount + 1;
    }
    if (keyCount == 1){ b := true; }
    else { b := false; }
    i := i + 1;
  }
  (*@\setlength{\fboxsep}{0pt}\colorbox{posHighlight}{\strut \color{white} // Retrieved problem-independent proof strategy hint}@*)
  (*@\setlength{\fboxsep}{0pt}\colorbox{posHighlight}{\strut\color{white} \textbf{assert} a[..] == a[..a.Length];} @*)
}
\end{lstlisting}
\end{minipage}
\caption{Problem-independent array slicing hint completes the proof.}
\label{fig:hintAug}
\end{figure}

Fig.~\ref{fig:hintAug} illustrates a third limitation: the generated invariant clauses may be sufficient to prove the post-condition, yet the Dafny verifier still requires additional hints (e.g., assertions) to complete the proof. Crucially, these hints are independent of the base program logic and instead capture structural properties of specific data types (e.g., array slicing).

This observation motivates constructing a small set of problem-independent proof strategies that can be retrieved at inference time and applied across problems. We demonstrate that only eight such strategies suffice to significantly improve performance (all reported in Appendix~\ref{appen:hintAugdetails}). While these hints are currently generated manually by an expert, we show in Appendix~\ref{append:llmhintgen} that their generation can be automated: an LLM can extract hints by comparing failed attempts with ground-truth solutions.

During inference, the hints and all failed attempts are provided to the LLM, which retrieves relevant parts for subsequent generation. Fig.~\ref{fig:hintAug} demonstrates that LLMs can successfully retrieve appropriate hints to establish the proof.

\section{\ToolName}

\subsection{Algorithm}

\begin{algorithm}[t]
\caption{\ToolName}
\label{alg:dafnypro}
\KwIn{Base Dafny program $P$ (without annotations, with specifications), LLM $\mathcal{L}$ for annotation generation, set of tactics $\mathcal{T}$, maximum attempts $N$}
\KwOut{Annotated Dafny program $P^\ast$ that verifies, or failure if not found}
$i \gets 0$\;
$\mathcal{R} \gets \emptyset$ \hfill\tcp{Relevant hints}
$\mathcal{P} \gets \emptyset$ \hfill\tcp{Store all attempts}

\While{$i < N$}{
    $P_i \gets \mathcal{L}(\mathcal{P}, \mathcal{R}, i)$; \hfill \tcp{Use info from all attempts}
    
    \If{\textbf{DiffChecker}$(P_i, P) = \text{fail}$}{
        \textbf{continue};\hfill \tcp{Check soundness}
    }    
    \If{\textbf{Verifier}$(P_i) = \text{success}$}{
        \Return $P_i$\;
    }
    \If{ \textbf{Verifier}$(P_i)$ reports non-inductive invariants}{
        $P_i' \gets \text{PruneNonInductive}(P_i)$\;
        
        \If{\textbf{Verifier}$(P_i') = \text{success}$}{
            \Return $P_i'$\;
        }
    }
    \tcp{Retrieve relevant tactics for $P_i$}
    $\mathcal{R} \gets \text{RetrieveTactics}(\mathcal{T}, P_i)$
    
    $\mathcal{P} \gets \mathcal{P} \bigcup \{P_i\}$
    
    $i \gets i + 1$\;
}
\Return failure\;
\end{algorithm}

\ToolName integrates the three mechanisms from Sec.~\ref{sec:techContribution}. The algorithm takes as input a base Dafny program with specifications, an LLM for annotation generation, and a fixed set of problem-independent hints. It attempts to produce a verified annotated program within a fixed number of attempts.

Fig.~\ref{fig:DafnyProPipeline} outlines the key steps, with pseudocode in Algorithm~\ref{alg:dafnypro}. \ToolName iteratively generates annotated versions of the base program using the LLM (line 5). For each generated program, the diff-checker verifies that annotations do not alter the original program logic, ensuring soundness (lines 6--7). The Dafny verifier then attempts to prove correctness with respect to the specifications. If non-inductive invariant clauses are detected, they are pruned and the program is re-verified (lines 11--13). Relevant hints are extracted from each attempt and incorporated into subsequent generations (line 14). This process continues until either verification succeeds or the attempt limit is reached.

\subsection{Evaluation}
\label{sec:dafnyproEval}

We evaluate \ToolName on four datasets: Clover \cite{clover}, MBPP-Dafny \cite{mbppDfy}, HumanEval-Dafny \cite{HumanEvalDafny2023}, and DafnyBench \cite{dafnybench}. Each dataset consists of unannotated base programs with specifications; the task is to generate annotations (e.g., loop invariants, assertions) necessary for verification to succeed. DafnyBench is the largest and most challenging, containing 782 programs with an average of 52.77 lines of code and a maximum of 41.3k lines (see Table~\ref{tab:benchmark_stats_basic} in the Appendix).

\paragraph{Dataset Quality}
As described earlier, several base programs in the existing DafnyBench dataset are not compilable. Although the annotated ground-truth programs scraped from the internet compile successfully, the ad hoc methods used to remove annotations introduce compilation errors (example in Appendix~\ref{append:dataCorrect}). Our diff-checker provides a principled approach to removing annotations from ground-truth programs. Using it, we fixed all 70 of 782 base programs that had compilation errors, improving the percentage of verified programs (Table~\ref{tab:model_comparison}).

\paragraph{Setup}
We evaluate \ToolName on Claude 3.5 Sonnet and Claude 3.7 Sonnet, both accessed through Amazon Bedrock. Program verification uses Dafny $4.11.0$. Experiments run on a machine with 16 CPU cores and 90 GB of RAM. We set the number of attempts to $N = 10$ for all experiments.

\paragraph{Results}

\begin{table}[t]
\centering
\begin{subtable}{0.48\textwidth}
\centering
\resizebox{\columnwidth}{!}{%
\begin{tabular}{l c c c c}
\hline
\textbf{Model} & \textbf{Clover} & \textbf{MBPP} & \textbf{HumanEval} & \textbf{DafnyBench} \\
\hline
Claude 3.5 Sonnet & 97.43 & 97.76 & 95.32 & 86.18 \\
Claude 3.7 Sonnet & 100.0 & 98.73 & 94.39 & 85.67 \\
\hline
\end{tabular}
}
\caption{Performance across benchmarks}
\label{tab:model_performance}
\end{subtable}
\hfill
\begin{subtable}{0.48\textwidth}
\centering
\resizebox{\columnwidth}{!}{%
\begin{tabular}{l l c c c}
\hline
\textbf{Model} & \textbf{Dataset} & \textbf{Benchmark Fixes} & \textbf{Pruning} & \textbf{Hint Augmentation} \\
\hline
Claude 3.5 Sonnet & HumanEval & X & +12.26\% & +33.64\% \\
Claude 3.7 Sonnet & HumanEval & X & +5.66\% & +13.08\% \\
Claude 3.5 Sonnet & DafnyBench & +3.58\% & +3.19\% & +9.97\% \\
Claude 3.7 Sonnet & DafnyBench & +3.96\% & +3.06\% & +10.86\% \\
\hline
\end{tabular}
}
\caption{Component contributions}
\label{tab:model_comparison}
\end{subtable}
\caption{\ToolName performance across different models and benchmarks.}
\label{tab:combined_results}
\end{table}

\begin{figure}[t]
\centering
\begin{subfigure}[t]{0.8\columnwidth}    
    \centering
    \includegraphics[width=\linewidth]{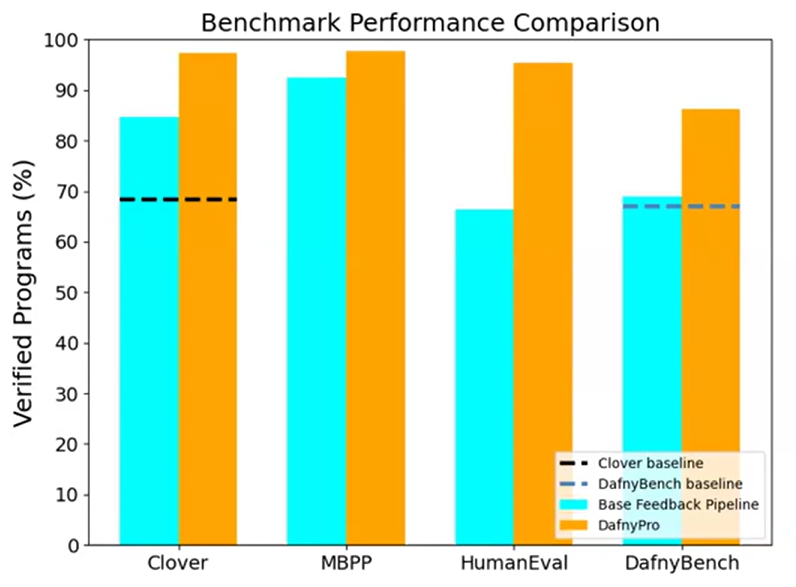}
    \caption{\ToolName performance with Claude 3.5 Sonnet across benchmarks.}
    \label{fig:claude3.5}
\end{subfigure}
\hfill
\begin{subfigure}[t]{0.8\columnwidth}
    \centering
    \includegraphics[width=\linewidth]{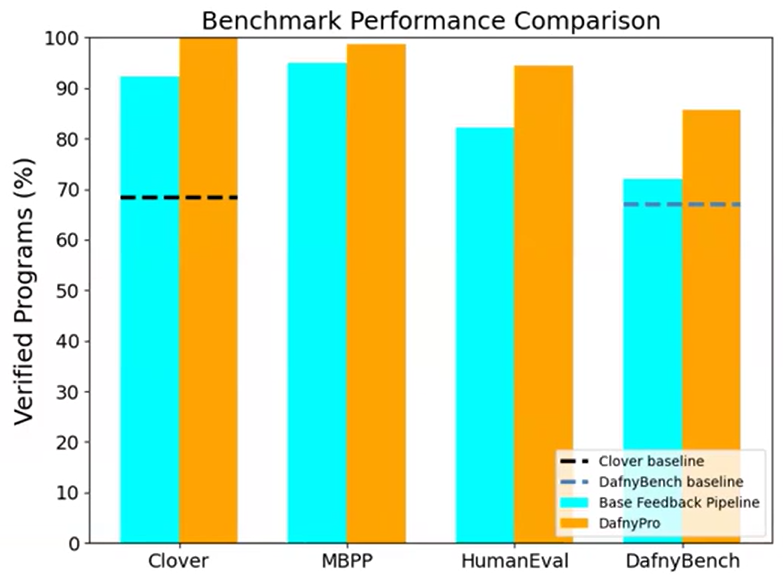}
    \caption{\ToolName performance with Claude 3.7 Sonnet across benchmarks.}
    \label{fig:claude3.7}
\end{subfigure}
\caption{\ToolName performance with Claude 3.5 and 3.7 Sonnet.
}
\label{fig:modelPerformance}
\end{figure}

\begin{figure}[t]
    \centering
\includegraphics[width=0.8\columnwidth]{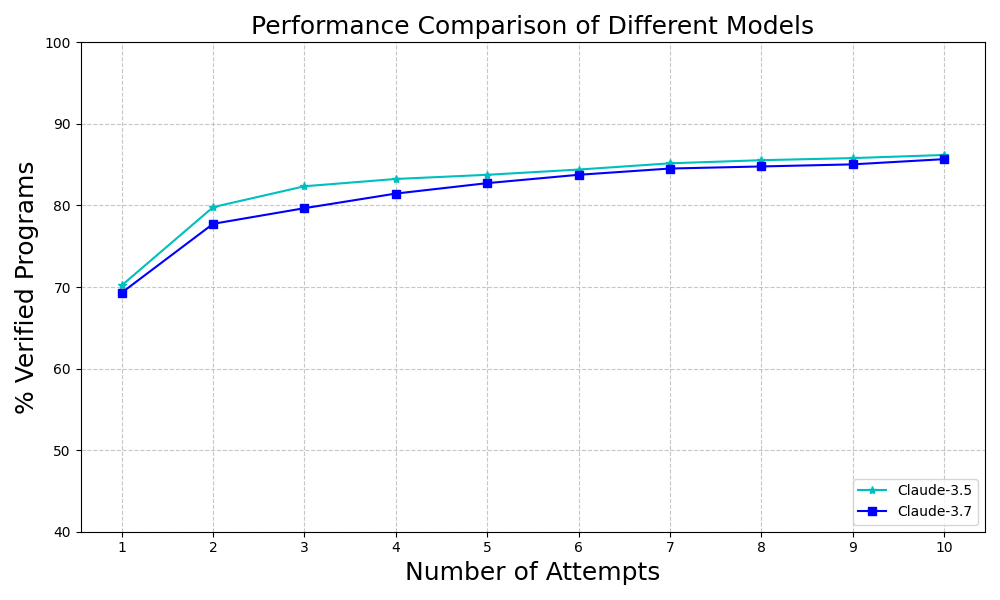}
    \caption{\ToolName verification rates across iterations on DafnyBench.}
\label{fig:iterPerformance}
\end{figure}

Table~\ref{tab:model_performance} summarizes \ToolName's performance across all four benchmarks. We report the percentage of base programs verified using LLM-generated annotations. Both Claude Sonnet models with \ToolName exceed $85$\% on the challenging DafnyBench dataset. Figs.~\ref{fig:claude3.5} and \ref{fig:claude3.7} show performance across all benchmarks. The baseline relies solely on Dafny verifier feedback to iteratively generate candidates, without pruning or hint augmentation. Dotted lines represent previously reported results from Clover \cite{clover} and DafnyBench \cite{dafnybench}. Lower verification rates in prior work stem from older models (e.g., Claude 3 Opus in DafnyBench \cite{dafnybench}) and overestimation due to "cheating." \ToolName achieves a $5$\% improvement on MBPP and $14$-$16$\% improvements on the other three datasets over the baseline for both Claude models.

Table~\ref{tab:model_comparison} provides a detailed breakdown showing that hint augmentation yields the largest improvements. The third column reports percentage gains from fixing the 70 DafnyBench programs, further highlighting the diff-checker's importance. 
Verified accuracy over all 10 attempts on DafnyBench with different models is shown in Fig.~\ref{fig:iterPerformance}.

\section{Fine-Tuned Models for Efficient Deployment}

\subsection{Data Curation}

In DafnyPro, the LLM responsible for annotation generation can be any capable model. Although large models like Claude achieve strong performance, their high computational cost motivates lightweight alternatives that can run efficiently. However, the limited availability of Dafny programs online poses a challenge for large-scale fine-tuning of smaller models. To address this, we leverage DafnyPro itself—powered by larger models—to curate high-quality training data.

This curated data serves two purposes: (1) training smaller models to generate correct annotations, and (2) teaching them to learn from Dafny verifier feedback. Including both successful and failed verification attempts allows models to interpret verifier feedback and correct errors in subsequent iterations. We collect data from all models in Table~\ref{tab:model_performance} running with \ToolName, pairing each incorrect attempt with its verified ground-truth annotations. Additionally, verifier error messages are automatically informalized, producing examples that teach the model to translate formal feedback into actionable guidance. Through this approach, a relatively small set of high-quality examples effectively fine-tunes smaller models, making them competitive with large LLMs while remaining computationally efficient. The total fine-tuning dataset contains 31k examples from all model attempts.

\subsection{Evaluation}

\begin{figure}[t]
    \centering
    \begin{subfigure}[t]{0.8\columnwidth}
        \centering
        \includegraphics[width=\textwidth]{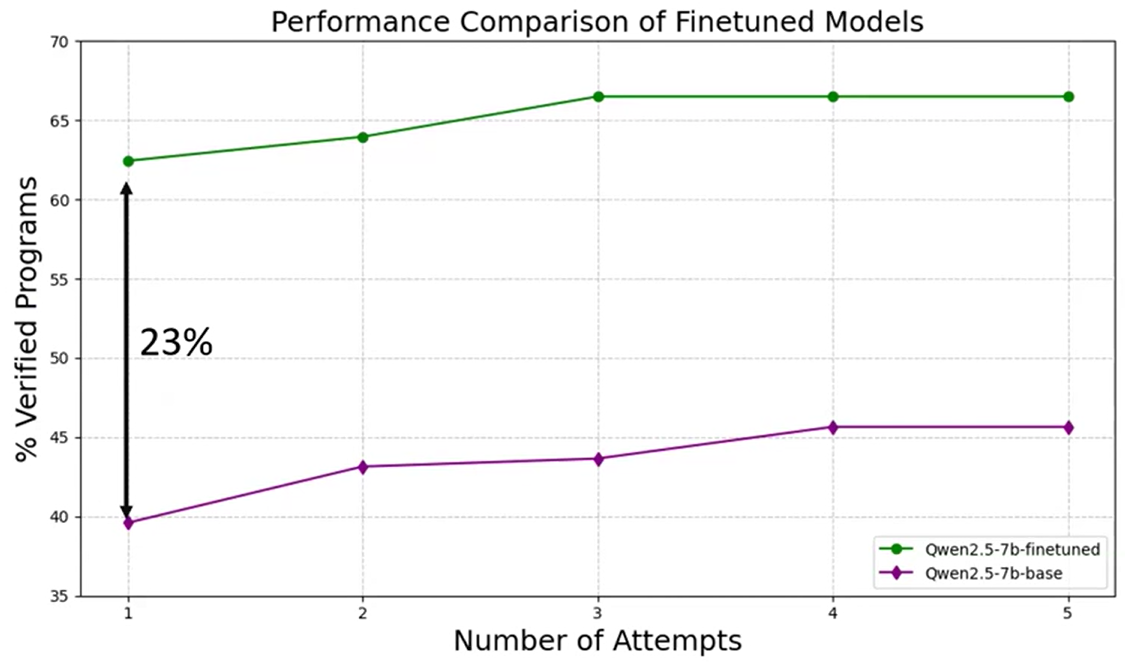}
        \caption{Qwen2.5-7B}
        \label{fig:1}
    \end{subfigure}
    \hfill
    \begin{subfigure}[t]{0.8\columnwidth}
        \centering\includegraphics[width=\textwidth]{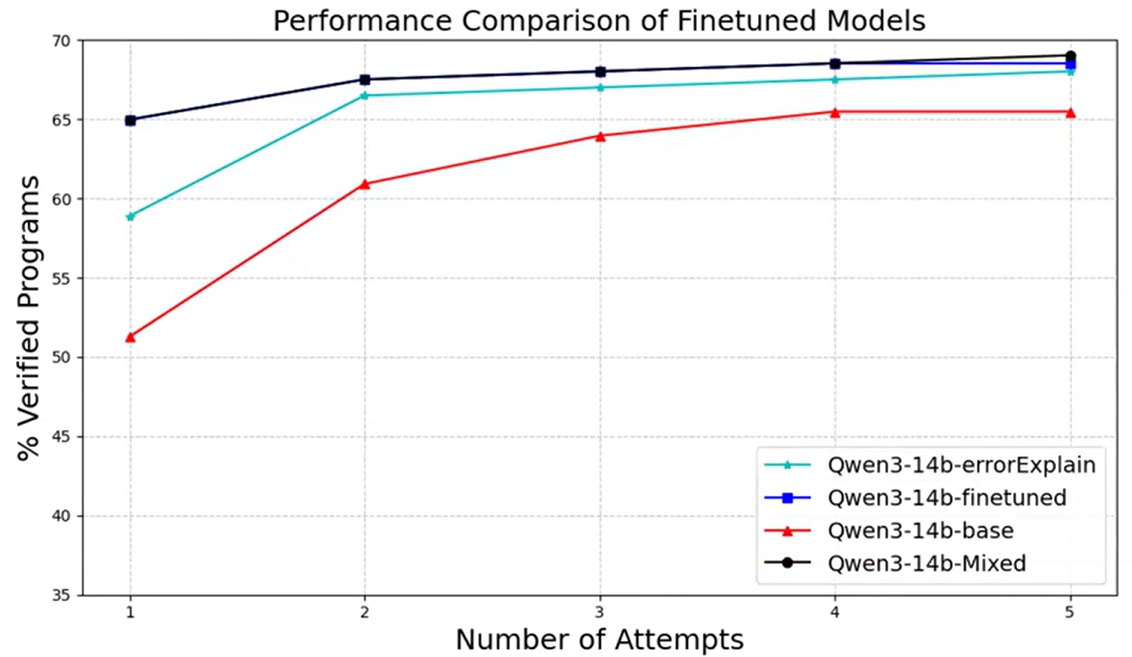}
        \caption{Qwen3-14B}
        \label{fig:3}
    \end{subfigure}
    \hfill
    \caption{Fine-tuning impact on DafnyBench performance.}
    \label{fig:four_figs}
\end{figure}

\paragraph{Setup}
We apply LoRA-based fine-tuning using the Hugging Face TRL library on Qwen2.5-7B and Qwen3-14B parameters, with rank $r=512$. All hyperparameters are given in Appendix~\ref{append:finetuneParams}. Evaluation uses a held-out test split of 197 programs from DafnyBench. Neither these programs nor any LLM-generated attempts on them were used during fine-tuning, preventing data leakage. For all base and fine-tuned models, we use \ToolName with $N=5$ attempts.

\paragraph{Results}
Fine-tuning substantially improves verification performance for both model sizes. On the first attempt, the fine-tuned Qwen2.5-7B model achieves a 23\% increase in verified programs compared to the base model (Fig.~\ref{fig:1}), and the fine-tuned Qwen3-14B model achieves a 13\% improvement over its base model (Fig.~\ref{fig:3}).

When fine-tuning the 14B model with both error explanations and ground-truth data, we observe slightly lower first-attempt performance compared to fine-tuning with ground-truth data only. However, after five attempts, the error-explanation fine-tuned model recovers and outperforms the ground-truth-only variant. This indicates that fine-tuning with error explanations enhances the model's ability to interpret and respond to Dafny verifier feedback during iterative refinement. To exploit this complementary behavior, we construct a mixing strategy: the first three attempts use the ground-truth fine-tuned model, and the final two use the error-explanation fine-tuned model. This achieves the highest verification rate of 69.5\% (Fig.~\ref{fig:3}), demonstrating that combining both fine-tuning paradigms yields superior performance. For comparison, the similarly sized fine-tuned LLaMA-3.1 8B model from \cite{dafnyAnnotator} achieves 50\% verified performance, which is surpassed by our Qwen2.5-7B model.
\section{Conclusion and Future Work}

We introduced \ToolName, an inference-time framework that achieves state-of-the-art verification accuracy across multiple datasets, including DafnyBench. By ensuring soundness through diff-checking, improving efficiency via pruning, and leveraging proof strategies through hint augmentation, \ToolName demonstrates substantial improvements over baselines. We also developed a suite of fine-tuned models that, when integrated with \ToolName, deliver competitive verification performance while maintaining computational efficiency. Together, these contributions enable practical formal reasoning systems powered by large language models.

Despite \ToolName's strong empirical performance, several opportunities exist to broaden its scope. First, our comparison among foundation models includes only Claude 3.5/3.7 Sonnet. Evaluating a broader range of recent models would provide a more comprehensive assessment of \ToolName's generalizability. Second, our fine-tuning experiments focus on supervised fine-tuning of Qwen models. While this strategy proved effective, incorporating reinforcement learning-based fine-tuning techniques, as explored in prior work~\cite{preface}, could further improve the model's ability to utilize verifier feedback and strengthen iterative reasoning. Future work will investigate these directions and test additional architectures to deepen understanding of the framework's robustness and scalability.

\clearpage
\newpage
\onecolumn
\bibliographystyle{ACM-Reference-Format}
\bibliography{dafnyPro}

\appendix
\clearpage
\newpage
\onecolumn
\section{Appendix}
\subsection{Diff-Checker Details}
\label{appen:diffCheck}

\begin{figure}[!htbp]
    \centering
    
    % Subfigure 1 
    \begin{subfigure}{0.96\columnwidth}
    \begin{minipage}{\linewidth}
    \begin{lstlisting}
b := a+1;  
while ( b < n )
// FILL IN INVARIANTS.
{
  if ( X[b] <= p ) {
    var t := X[b];
    X[b] := X[a];
    X[a] := t;
    a := a + 1;
  }
  b := b + 1;
}
    \end{lstlisting}
    \end{minipage}
    \caption{Base code without annotations.}
    \end{subfigure}

    \vspace{0.4em}

    % Subfigure 2 
    \begin{subfigure}{0.96\columnwidth}
    \begin{minipage}{\linewidth}
    \begin{lstlisting}
(*@\setlength{\fboxsep}{0pt}\colorbox{highlightred}{\strut \color{white} b := a;  // Base code altered (was: b := a+1)}@*)
while ( b < n )
  invariant a <= b <= n
{
  if ( X[b] <= p ) {
    var t := X[b];
    X[b] := X[a];
    X[a] := t;
    a := a + 1;
  }
  b := b + 1;
}
    \end{lstlisting}
    \end{minipage}
    \caption{LLM modifies base code instead of only adding annotations, compromising soundness.}
    \end{subfigure}

    \vspace{0.4em}

    % Subfigure 3 
    \begin{subfigure}{0.96\columnwidth}
    \begin{minipage}{\linewidth}
    \begin{lstlisting}
b := a+1;  
while ( b < n )
invariant 0 <= a < b <= n+1
invariant (b == n+1) ==> (a == n)
{
  if ( X[b] <= p ) {
    var t := X[b];
    X[b] := X[a];
    X[a] := t;
    a := a + 1;
  }
  b := b + 1;
}
    \end{lstlisting}
    \end{minipage}
    \caption{Ground truth solution.}
    \end{subfigure}

    \caption{Examples of LLM-induced modifications in base code for cheating.}
    \label{fig:exCheatingFull}
\end{figure}

\clearpage
\newpage 
\subsection{Pruning Details}
\label{appen:pruningDetails}
\begin{figure}[h]
\centering

\begin{minipage}{\linewidth}
\begin{lstlisting}
method onlineMax(a: array<int>, x: int) returns (ghost m: int, p: int)
  requires 1 <= x < a.Length
  requires a.Length != 0
  ensures x <= p < a.Length
  ensures forall i :: 0 <= i < x ==> a[i] <= m
  ensures exists i :: 0 <= i < x && a[i] == m
  ensures x <= p < a.Length-1 ==> (forall i :: 0 <= i < p ==> a[i] < a[p])
  ensures (forall i :: x <= i < a.Length && a[i] <= m) ==> p == a.Length - 1
{
  p := 0; var best := a[0]; var i := 1; i := x;
  // ... Code Block
  while i < a.Length
    invariant x <= i <= a.Length
    invariant forall j :: 0 <= j < x ==> a[j] <= m
    invariant exists j :: 0 <= j < x && a[j] == m
    invariant forall j :: x <= j < i ==> a[j] <= m
    invariant i < a.Length ==> p == 0
    (*@\setlength{\fboxsep}{0pt}\colorbox{posHighlight}{\strut \color{white} // Removed: non-inductive and unnecessary for proving post-condition
} @*)
    (*@\setlength{\fboxsep}{0pt}\colorbox{posHighlight}{\strut \color{white} invariant i == a.Length ==> p == a.Length - 1} @*) 
  {
    if a[i] > best {
      p := i;
      return;
    }
    i := i + 1;
  }
  p := a.Length - 1;
}
\end{lstlisting}
\end{minipage}
\caption{Pruning unnecessary invariants enables verification, avoiding wasted feedback iterations.}
\label{fig:pruningBenefitFulll}
\end{figure}

\clearpage
\newpage
\subsection{Hint Augmentation Details}
\label{appen:hintAugdetails}
\subsection*{Tactic: Bridging Partial and Full Array Slices}

When working with array slices in loop invariants, add explicit assertions to connect the final loop state with the postcondition's array representation.

\textbf{Specifically:}
\begin{itemize}
  \item For an array \texttt{a}, if your loop invariant uses \texttt{a[..i]} and your postcondition uses \texttt{a[..]}
  \item Add \texttt{assert a[..a.Length] == a[..];} after the loop
  \item This bridges the gap between the loop's final state (\texttt{i == a.Length}) and the postcondition's requirements
\end{itemize}

This pattern applies to any array verification problem where:
\begin{enumerate}
  \item Loop invariants track properties over \texttt{a[..i]}
  \item Postconditions reference properties over \texttt{a[..]}
  \item The connection between these two slice representations needs to be made explicit for the verifier
\end{enumerate}

\subsection*{Tactic: Connecting Loop-Local Sequences to Final Arrays}

When loop invariants track properties over partial array slices but postconditions reference complete arrays, add explicit assertions to bridge the semantic gap between loop-local reasoning and method-level guarantees.

\textbf{Specifically:}
\begin{itemize}
  \item If your loop invariant uses \texttt{arr[..i]} and your postcondition uses \texttt{arr[..]}
  \item If your loop builds a sequence that becomes an array in the postcondition
  \item Add assertions after the loop to connect the sequence representation with the final array representation
  \item This establishes the equivalence between the loop's accumulated result and the method's return value
\end{itemize}

This pattern applies to any array construction verification problem where:
\begin{enumerate}
  \item Loop invariants track properties over partial input and build intermediate collections
  \item Postconditions reference properties over complete input and final output arrays
  \item The verifier needs an explicit connection between intermediate sequences and final arrays
  \item The semantic relationship between loop-accumulated data and method postconditions requires bridging
\end{enumerate}

\emph{Key insight:} Dafny may not automatically recognize that a sequence used in loop reasoning corresponds to an array slice in the postcondition, even when they contain identical elements. Explicit assertions make this relationship visible to the verifier.

\subsection*{Tactic: Assisting Recursive Reasoning over Growing Array Slices}

When loop invariants involve recursive function calls on array slices, add explicit assertions to help the verifier understand the relationship between consecutive slice states and the final postcondition.

\textbf{Specifically:}
\begin{itemize}
  \item If your invariant uses a recursive function on \texttt{arr[..i]} and the postcondition uses the same function on \texttt{arr[..]}
  \item Add \texttt{assert arr[..i+1] == arr[..i] + [arr[i]];} inside the loop
  \item Add \texttt{assert arr[..] == arr[..i];} after the loop
  \item These assertions expose how recursive functions behave on incrementally growing slices
\end{itemize}

This pattern applies when:
\begin{enumerate}
  \item Loop invariants track recursive properties over \texttt{arr[..i]}
  \item Postconditions reference the recursive function over \texttt{arr[..]}
  \item The function's behavior under slice concatenation must be made explicit
  \item Incremental progress must align with the recursive definition
\end{enumerate}

\emph{Key insight:} While relationships like \texttt{Sum(arr[..i+1]) = Sum(arr[..i]) + arr[i]} are obvious to humans, Dafny benefits from explicit assertions that decompose slice growth into verifier-friendly steps.

\subsection*{Tactic: Simplifying Overly Complex Loop Invariants}

When loop invariants become overly complex due to slice-based membership reasoning, simplify them by strengthening invariants to use global properties.

\textbf{Specifically:}
\begin{itemize}
  \item Replace complex conditions tracking membership in \texttt{values[..idx]} with a stronger global invariant
  \item Remove redundant bounds like \texttt{0 <= idx} when implied elsewhere
  \item Retain essential invariants that directly support the postcondition
\end{itemize}

This applies when:
\begin{enumerate}
  \item Invariants track membership in growing slices
  \item Edge cases require complex conditional logic
  \item Postconditions depend on global properties
  \item Over-specification increases verification difficulty without improving correctness
\end{enumerate}

\emph{Key insight:} Stronger global invariants are often easier to verify than weaker, slice-local invariants that require extensive case analysis.

\subsection*{Tactic: Lexicographic Decreases for Mixed-Sign Recursion}

When recursive methods handle both positive and negative cases, use lexicographic decreases clauses to establish proper termination.

\textbf{Specifically:}
\begin{itemize}
  \item Replace absolute-value decreases with lexicographic clauses like \texttt{decreases x < 0, x}
  \item Prioritize negative cases before positive ones
  \item Ensure progress both in sign normalization and magnitude reduction
\end{itemize}

This pattern applies when:
\begin{enumerate}
  \item Recursion treats positive and negative inputs differently
  \item Negative inputs are converted to positive ones
  \item Simple decreases clauses fail to capture termination structure
\end{enumerate}

\emph{Key insight:} Lexicographic decreases communicate that sign normalization is progress, followed by numeric descent toward termination.

\subsection*{Tactic: Verifying Loop Invariant Initialization}

When loop invariants fail, check that they hold at loop entry by validating variable initialization.

\textbf{Specifically:}
\begin{itemize}
  \item Ensure initialized variables satisfy invariants before the first iteration
  \item Watch for off-by-one errors in index-based invariants
  \item Align initial values with function-based invariant definitions
\end{itemize}

\emph{Key insight:} Invariants must hold before, during, and after the loop. Many failures arise from invariants that are correct mid-loop but false at entry.

\subsection*{Tactic: Making Set Comprehension Updates Explicit}

When loops modify sets and track cardinalities via set comprehensions, add assertions that explain how comprehensions change.

This applies when:
\begin{enumerate}
  \item Invariants track sizes of filtered sets
  \item Elements move between sets
  \item Counters depend on element properties
  \item Dafny needs guidance on comprehension semantics
\end{enumerate}

\emph{Key insight:} Explicit assertions show whether adding an element expands a filtered set or leaves it unchanged, making cardinality updates verifiable.

\subsection*{Tactic: Conditional Invariants for Boundary Cases}

When strict range invariants fail in edge cases, replace them with conditional invariants.

This pattern applies when:
\begin{enumerate}
  \item Loops may not execute due to initial conditions
  \item Variables are uninitialized in boundary cases
  \item Strict bounds fail for inputs like \texttt{n = 0} or \texttt{n = 1}
\end{enumerate}

\emph{Key insight:} Conditional invariants of the form \texttt{edge\_case || normal\_case} ensure correctness at loop entry while preserving strong guarantees during normal execution.

\subsection{Benchmark Details}
\label{sec:benchDetails}
\begin{table}[h]
\centering
\caption{Benchmark Statistics}
\begin{tabular}{|l|r|r|r|r|l|}
\hline
\textbf{Benchmark} & \textbf{\# Samples with loops} & \textbf{\# Samples} & \textbf{Avg. LoC} & \textbf{Max LoC} & \textbf{Remarks} \\
\hline
Clover & 39 & 63 & 18.62 & 745 & Textbook Problems \\
\hline
MBPP-Dafny & 79 & 164 & 19.47 & 1558 & Converted from MBPP \\
\hline
HumanEval-Dafny & 107 & 132 & 50.45 & 5449 & Converted from HumanEval \\
\hline
DafnyBench & 600 & 782 & 52.77 & 41375 & Scraped from the internet \\
\hline
\end{tabular}
\label{tab:benchmark_stats_basic}
\end{table}
\clearpage
\newpage
\subsection{LLM Usage for Hint Generation}
\label{append:llmhintgen}
\begin{figure}[h]
\centering

\begin{minipage}{\linewidth}
\begin{lstlisting}
method only_once<T(==)>(a: array<T>, key: T) returns (b:bool)
ensures (multiset(a[..])[key] ==1 ) <==> b
{
var i := 0; b := false; var keyCount := 0;
while i < a.Length
 invariant 0 <= i <= a.Length
 invariant keyCount == multiset(a[..i])[key]
 invariant b <==> keyCount == 1
{
if (a[i] == key){
 keyCount := keyCount + 1;
}
if (keyCount == 1) { b := true; }
else { b := false; }
 i := i + 1;
}
}

\end{lstlisting}
\end{minipage}
\caption{The failed attempt that could not be verified.}
\label{fig:llmfailed}
\end{figure}

\begin{figure}[h]
\centering

\begin{minipage}{\linewidth}
\begin{lstlisting}
method only_once<T(==)>(a: array<T>, key: T) returns (b:bool)
ensures (multiset(a[..])[key] ==1 ) <==> b
{
var i := 0; b := false; var keyCount := 0;
while i < a.Length
 invariant 0 <= i <= a.Length
 invariant keyCount == multiset(a[..i])[key]
 invariant b <==> keyCount == 1
{
if (a[i] == key){
 keyCount := keyCount + 1;
}
if (keyCount == 1) { b := true; }
else { b := false; }
 i := i + 1;
}
assert a[..] == a[..a.Length];
}

\end{lstlisting}
\end{minipage}
\caption{The verified ground truth annotated code.}
\label{fig:llmver}
\end{figure}

\subsection*{LLM Generated Hint Based on Examples From Fig.~\ref{fig:llmfailed} and Fig.~\ref{fig:llmver}}
\subsection*{Tactic: Bridging Partial and Full Array Slices}

When working with array slices in loop invariants, add explicit assertions to connect the final loop state with the postcondition's array representation.

\textbf{Specifically:}
\begin{itemize}
  \item For an array \texttt{a}, if your loop invariant uses \texttt{a[..i]} and your postcondition uses \texttt{a[..]}
  \item Add \texttt{assert a[..a.Length] == a[..];} after the loop
  \item This bridges the gap between the loop's final state (\texttt{i == a.Length}) and the postcondition's requirements
\end{itemize}

This pattern applies to any array verification problem where:
\begin{enumerate}
  \item Loop invariants track properties over \texttt{a[..i]}
  \item Postconditions reference properties over \texttt{a[..]}
  \item The connection between these two slice representations needs to be made explicit for the verifier
\end{enumerate}

\clearpage
\newpage

\subsection{Dataset Correction}
\label{append:dataCorrect}

\begin{figure}[!htbp]
    \centering
    % Subfigure 2 
    \begin{subfigure}{0.96\columnwidth}
    \begin{minipage}{\linewidth}
    \begin{lstlisting}
method GetEven(s: array<nat>) modifies s
  ensures forall i :: 0 <= i < s.Length ==>
    if old(s[i]) % 2 == 1 then s[i] == old(s[i]) + 1
    else s[i] == old(s[i])
{
  var i := 0;
  while i < s.Length
    invariant 0 <= i <= s.Length
    invariant forall j :: 0 <= j < i ==>
      if old(s[j]) % 2 == 1 then s[j] == old(s[j]) + 1
      else s[j] == old(s[j])
    invariant forall j :: i <= j < s.Length ==> s[j] == old(s[j])
  {
    if s[i] % 2 == 1 {
      s[i] := s[i] + 1;
    }
    i := i + 1;
  }
}

    \end{lstlisting}
    \end{minipage}
    \caption{Ground truth code with multiline invariant.}
    \end{subfigure}

    \vspace{0.4em}

    % Subfigure 3 
    \begin{subfigure}{0.96\columnwidth}
    \begin{minipage}{\linewidth}
    \begin{lstlisting}
method GetEven(s: array<nat>) modifies s
  ensures forall i :: 0 <= i < s.Length ==>
    if old(s[i]) % 2 == 1 then s[i] == old(s[i]) + 1
    else s[i] == old(s[i])
{
  var i := 0;
  while i < s.Length

    (*@\setlength{\fboxsep}{0pt}\colorbox{highlightred}{\strut \color{white}   if old(s[j]) \% 2 == 1 then s[j] == old(s[j]) + 1   // Compilation error due to partially removed invariant} @*)
    (*@\setlength{\fboxsep}{0pt}\colorbox{highlightred}{\strut \color{white}  else s[j] == old(s[j])} @*)
    
  {
    if s[i] % 2 == 1 {
      s[i] := s[i] + 1;
    }
    i := i + 1;
  }
}

    \end{lstlisting}
    \end{minipage}
    \caption{Compilation error introduced by adhoc annotation remover.}
    \end{subfigure}

    \caption{Examples of compilation errors in base code in DafnyBench.}
    \label{fig:dataCorrect}
\end{figure}

% \begin{lstlisting}[language=Dafny, caption={GetEven method in Dafny}, label={lst:geteven}]
% \end{lstlisting}

\clearpage
\newpage
\subsection{Finetuning Hyperparameters}
\label{append:finetuneParams}
{
\lstset{
  language=Python,
  basicstyle=\ttfamily\footnotesize,
  keywordstyle=\color{blue},
  stringstyle=\color{teal},
  commentstyle=\color{gray},
  showstringspaces=false,
  frame=single,
  rulecolor=\color{gray},
  breaklines=true,
  aboveskip=0.5em,
  belowskip=0.5em,
  columns=fullflexible,
  lineskip=-1pt
}
\begin{lstlisting}[caption={LORA finetuning config used for all models.}, label={lst:perfConfing}]
"peft_config": {
    "value": {
      "default": {
        "r": 512,
        "bias": "none",
        "revision": null,
        "use_dora": false,
        "lora_bias": false,
        "peft_type": "LORA",
        "task_type": "CAUSAL_LM",
        "eva_config": null,
        "lora_alpha": 512,
        "use_qalora": false,
        "use_rslora": true,
        "auto_mapping": null,
        "corda_config": null,
        "lora_dropout": 0.05,
        "megatron_core": "megatron.core",
        "fan_in_fan_out": false,
        "inference_mode": false,
        "layers_pattern": null,
        "runtime_config": {
          "ephemeral_gpu_offload": false
        },
        "target_modules": [
          "v_proj",
          "q_proj"
        ],
        "exclude_modules": null,
        "megatron_config": null,
        "modules_to_save": null,
        "init_lora_weights": true,
        "layer_replication": null,
        "qalora_group_size": 16,
        "layers_to_transform": null,
        "trainable_token_indices": null
      }
    }
  }
\end{lstlisting} 
}
\end{document}